\documentclass[aps,prl,twocolumn,groupedaddress,floatfix,showpacs]{revtex4-1}
\usepackage{amssymb}
\usepackage{graphicx}
\usepackage{subfigure}	 	
\usepackage[english]{babel}
\usepackage{float}
\usepackage{color}

\newcommand{\be}{\begin{equation}}
\newcommand{\ee}{\end{equation}}

\begin{document}
\title{Nonlinear light localization around the core of a `holey' fiber}

\author{Francis H. Bennet$^1$ and Mario I. Molina$^{2,3}$}

\affiliation{$^1$Nonlinear Physics Centre, Research School of Physics and Engineering, Australian National University, 0200 ACT, Canberra, Australia \\
$^2$Departmento de F\'{\i}sica, Facultad de Ciencias, Universidad de Chile, Santiago, Chile\\
$^3$Center for Optics and Photonics, Universidad de Concepci\'{o}n, Concepci\'{o}n, Chile}

\begin{abstract}
We examine localized surface modes in the core of a photonic crystal fiber composed of a finite nonlinear (Kerr)  hexagonal waveguide array with a central defect. Using a discrete approach, we find the fundamental surface mode and its stability window. We also examine an
unstaggered, ring-shaped surface mode and find that it is always unstable, decaying to the single-site fundamental surface mode. A continuous model computation reveals that an initial vortex excitation ($S=1$) of small amplitude around the central hole can survive for a relatively long evolution distance. At high amplitudes, however, it decays to a ring configuration with no well-defined phase structure.
\end{abstract}

\pacs{42.81.Qb, 05.45.Yv, 42.65.Tg, 42.65.Wi}
\maketitle

\section{Introduction}

Nonlinear propagation of fundamental Gaussian optical beams has produced a rich variety of physical phenomena such as discrete and gap solitons in positive and negative periodic nonlinear media \cite{Christodoulides1988,Eisenberg1998,Kivshar1993}. We can uncover an even wider range of novel nonlinear optical propagation, by studying modes with different symmetries. One such mode is an optical vortex, which is an optical mode including a phase singularity at its centre.

Optical vortices and their propagation have been studied for their ability to trap and manipulate particles \cite{Gahagan1999}, and in the production of waveguides in atomic vapor \cite{Truscott1999}. The nonlinear propagation of vortex modes in the core of a photonic crystal fiber (PCF) have been studied theoretically \cite{Ferrando2004, Johansson1998, Kevrekidis2001, Malomed2001, Kevrekidis2002}, and there have been theoretical and experimental study of vortex solitons in optically induced lattices \cite{Yang2003,Alexander2007,Neshev2004}. 

We take a new approach to the study of optical vortex propagation. Using an hexagonal array of nonlinear waveguides surrounding a solid core, we propagate an optical vortex in the waveguides adjacent to the core. We theoretically study the nonlinear propagation of vortex modes in this system, using both discrete and continuous models. 

Such structure is analogous to a liquid infiltrated photonic crystal fiber (PCF), which have been used to study nonlocal gap solitons \cite{Rasmussen2009}, the crossover from focusing to defocusing in a periodic array \cite{Bennet2010}, as well as the possibility for selective infiltration for a range of interesting structures and applications \cite{Bennet2010a,Wu2009,Vieweg2010}.

By propagating a vortex mode in waveguides around the solid core of a PCF we can study vortex modes interacting with a surface, where the periodic structure meets a homogenous dielectric. Such states have been studied and observed in similar structures for single site excitation with Gaussian modes \cite{Szameit2009,Szameit2008}.

This paper is organized as follows: Section II introduces the discrete model for an infiltrated PCF structure with an hexagonal geometry and a central defect making a solid core, in section III we introduce the continuous model of the same structure, focussing on the dynamical evolution of vortex excitations and finally, section IV concludes the paper. 

\section{Discrete model}

\begin{figure}[t]
\centerline{\includegraphics[width=1\columnwidth]{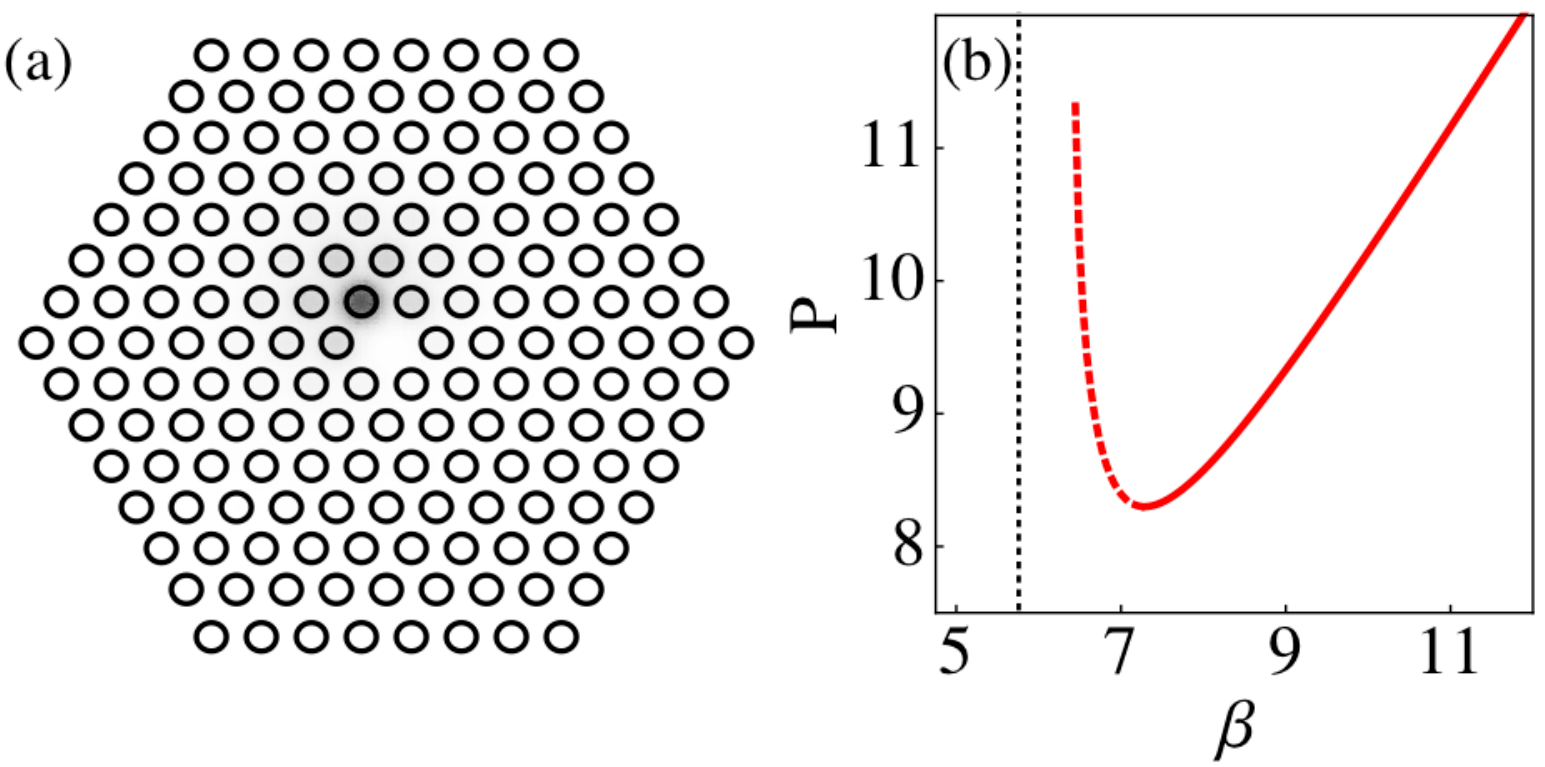}}
\caption{(Color online) Hexagonal waveguide array with central hole. (a) Example of a fundamental inner surface mode ($\beta=7.5, P=8.33$). Amount of shading denotes the distribution of optical intensity. (b) Power vs. propagation constant curve for this kind of mode. Solid (dashed) curve denotes stable (unstable) portions.}
\label{fig1}
\end{figure}

We consider a finite two-dimensional array of weakly-coupled nonlinear (Kerr) waveguides with hexagonal  geometry, with a missing waveguide at its center (Fig. \ref{fig1} (a)). In the framework of coupled-modes theory, the electric field $E(x,y,z)$ is presented as a superposition of (single) transverse modes $\phi(x,y)$ with amplitudes $U(z)$ that vary slowly along the longitudinal direction: $E(\vec{r},z)=\sum U_{\vec{n}}(z) \phi(\vec{r}-\vec{n})$, where $\vec{r}=(x,y)$ and $\vec{n}=(n_{x}, n_{y})$. These amplitudes obey the discrete nonlinear Schr\'{o}dinger equation,
\be
i {d U_{\vec{n}}\over{d z}} + V \sum_{\vec{m}\neq \vec{n}} U_{\vec{m}} + \gamma |U_{\vec{n}}|^{2} U_{\vec{n}}=0\label{eq:1}
\ee
where the sum is restricted to nearest-neighbors. The stationary solutions of Eq.(\ref{eq:1})
have the form $U_{\vec{n}}(z) = U_{\vec{n}} \exp(i \beta z)$, where $U_{\vec{n}}$ obeys
\be
-\beta\ U_{\vec{n}} + V \sum_{\vec{m}\neq \vec{n}} U_{\vec{m}} + \gamma |U_{\vec{n}}|^{2} U_{\vec{n}}=0.\label{eq:2}
\ee

\begin{figure}[h]
\centerline{\includegraphics[width=1\columnwidth]{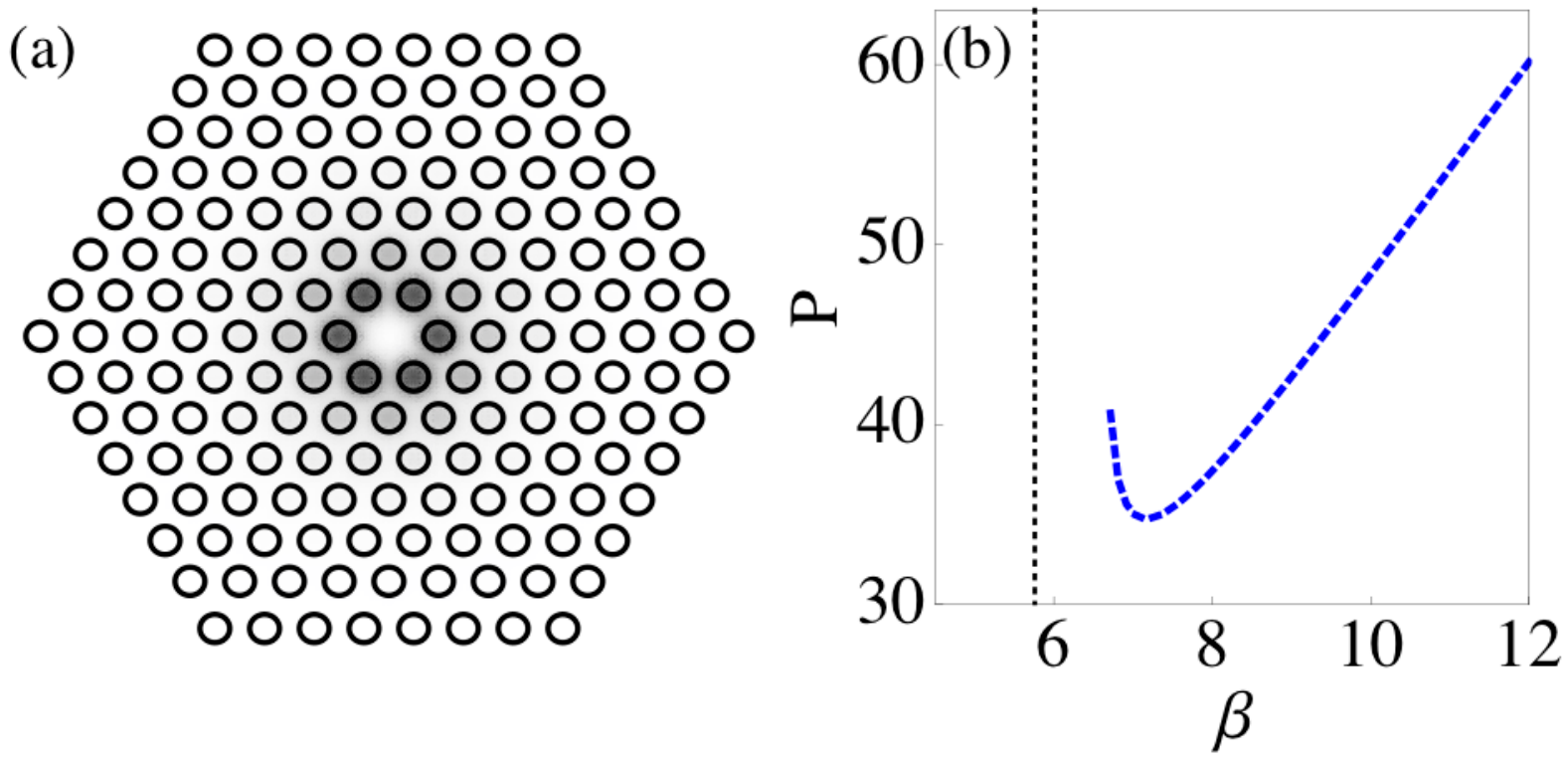}}
\caption{(Color online) Hexagonal waveguide array with central hole. (a) Example of an unstaggered `ring' surface mode ($\beta=8.0, P=37.4$).
Amount of shading denotes the distribution of optical intensity. (b) Power vs. propagation constant curve for this kind of mode. This mode is always linearly unstable.}
\label{fig2}
\end{figure}

\begin{figure}[h]
\centerline{\includegraphics[width=1\columnwidth]{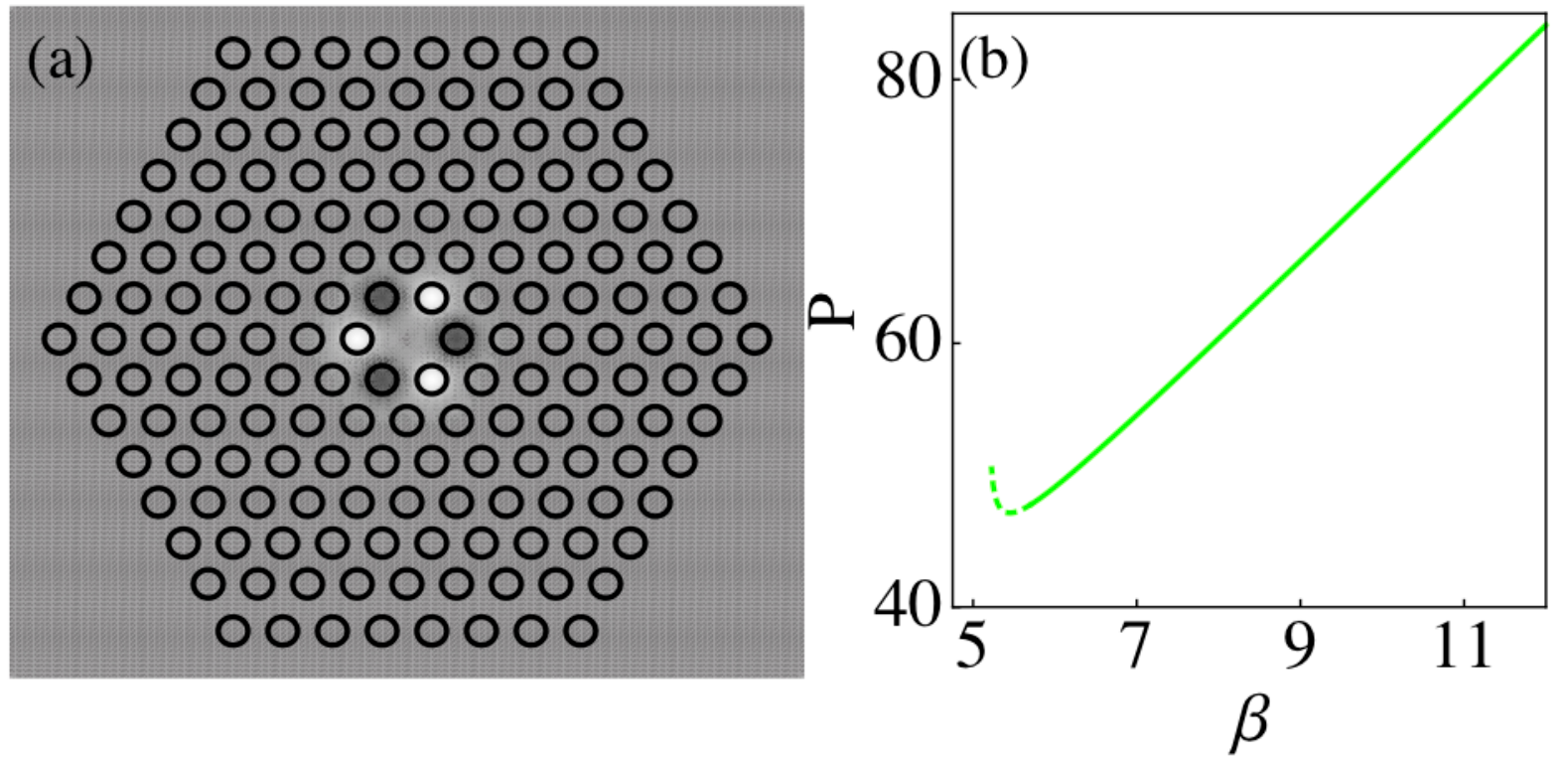}}
\caption{(Color online) Hexagonal waveguide array with central hole. (a) Example of an staggered `ring' surface mode ($\beta=5.75, P=47.95$).
Amount of shading denotes the distribution of the mode amplitude, going from white (amplitude=$-3$) to dark gray (amplitude=$+3$). (b) Power vs. propagation constant curve for this kind of mode. Solid (dashed) curve denotes stable (unstable) portions.}
\label{fig3}
\end{figure}

We are interested in localized modes centered around the boundary of the solid core of the array. The simplest of these `surface' modes is one centered on any of the six equivalent sites surrounding the missing guide. It is found by solving Eq. (\ref{eq:2}) using a direct extension of the Newton-Raphson method, starting from the decoupled (high-amplitude) limit, and performing a continuation process towards finite coupling values. For each mode found, we perform a standard linear stability analysis. Fig. \ref{fig1}(a) shows an  example of a spatial profile for this kind of mode, along with its power content $P=\sum |U_{n}|^{2}$ vs propagation constant $\beta$ curve (Fig. \ref{fig1}(b)). The curve obtained is typical of surface modes\cite{Molina2006} and obeys the Vakhitov-Kolokolov stability criterion. In order to approach something resembling a vortex-like mode, we consider next a higher-order mode, in the form of an unstaggered `ring' around the `hole', with no phase difference (i.e., zero vorticity). An example of this high-power mode is shown in Fig. \ref{fig2} along with its power vs. propagation constant curve. In this case, the mode is unstable for all values of its propagation constant. In fact, we find that most higher-order surface mode configurations are indeed unstable, with  the exception of one: The staggered version of the ring mode (Fig. \ref{fig3}(a)), where all amplitudes around the hole are identical initially, but with a phase difference of $\pi$ between nearest-neighbors around the ring. In this case, the mode is stable for initial amplitudes (Fig. \ref{fig3}(b)) exceeding a given threshold.

\begin{figure}[htb]
\includegraphics[width=1\columnwidth]{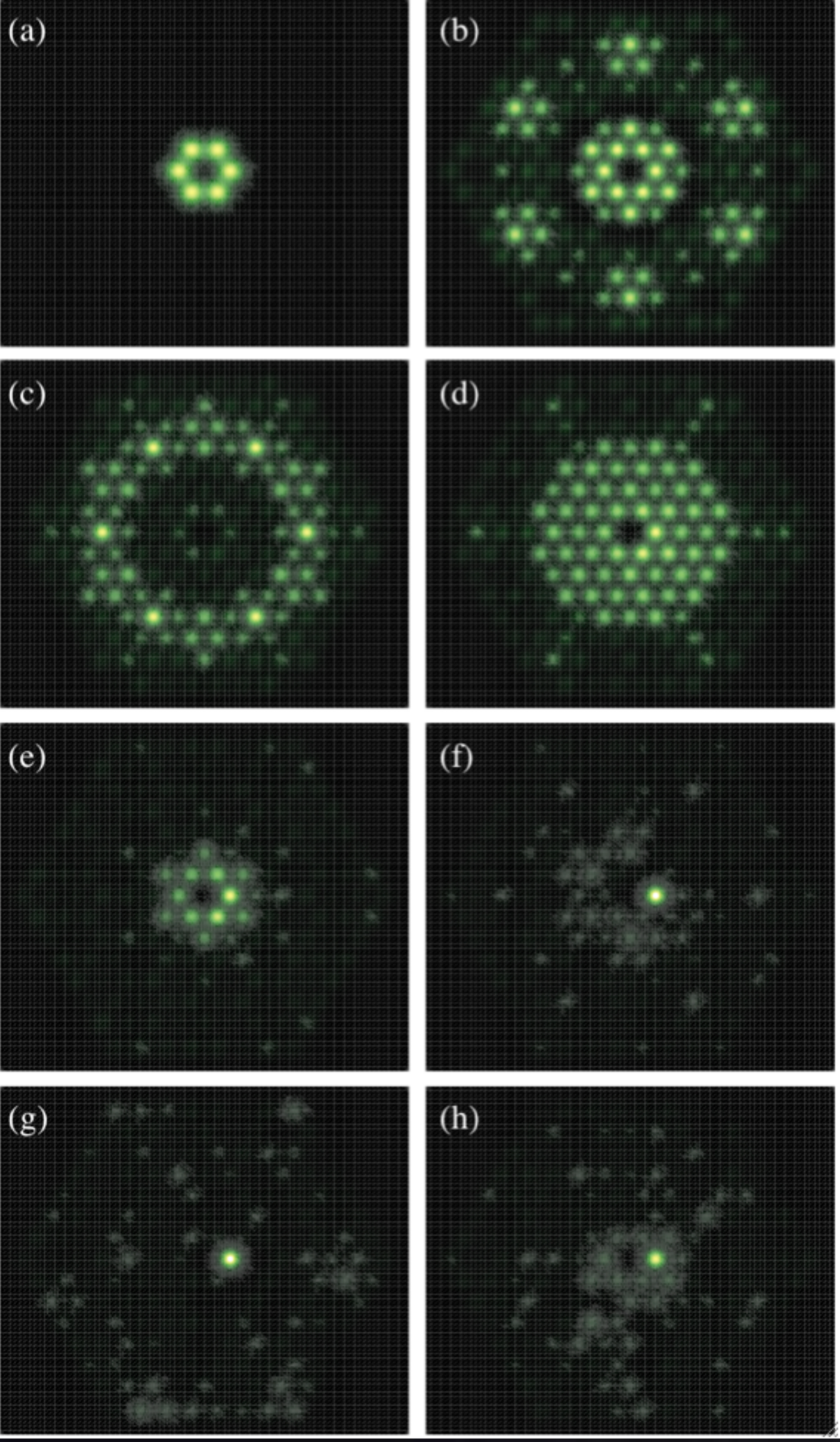}
\caption{(Color online) Evolution of unstable ring mode configuration over large propagation distance (a) $z=0$, (b) $z=50$, (c) $z=100$, (d) $z=136$, (e) $z=137$, (f) $z=138$(g) $z=150$, (h) $z=200$. The initial amplitude in all six sites around the `hole' is $2$, with no phase differences.}
\label{fig4}
\end{figure}

\begin{figure}[htb]
\includegraphics[width=7cm]{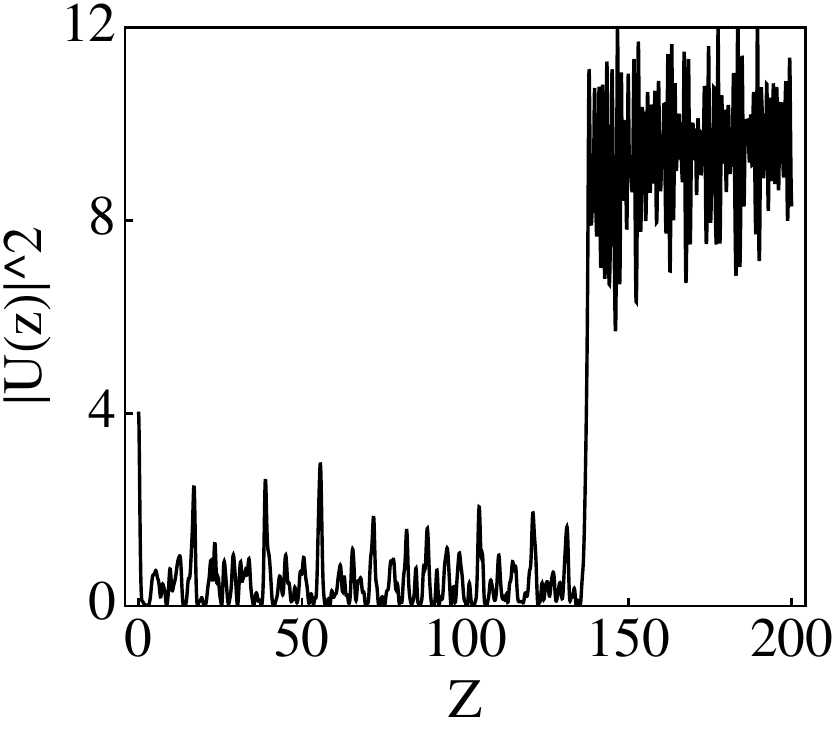}
\caption{(Color online) Evolution of unstable ring mode configuration over large propagation distance: Power content vs. longitudinal evolution distance, at the position of the center of the eventually self-trapped beam}
\label{fig5}
\end{figure}

One interesting question at this points is: If we excite dynamically the unstaggered (i.e., unstable)
ring configuration, what are the decay channels for this mode? Will it transition to the low-power, single-site stable mode, or will it change into the staggered (stable) ring, or perhaps  it will dissipate as radiation? To look for an answer, we follow the dynamical evolution of an initially completely localized ring mode configuration: $U_{\vec{n}}=U_{0}$ around the six sites surrounding the missing guide, $U_{\vec{n}}=0$ otherwise. Long-time evolution of this mode over large propagation distance for a finite sample of $N=168$ sites is shown in Fig. \ref{fig4}. Clearly, after a long transient behavior, where the diffracted beam bounces several times from the boundaries of the array, the beam becomes eventually selftrapped in one of the six possible fundamental mode configurations. It is interesting to note that this selftrapping transition is quite abrupt, as evidenced in Fig. \ref{fig5}. 

\section{Continuous model}
Given the geometry of the array, it is conceivable that the addition of vorticity could stabilize this (unstable) ring mode. After all, we know that in two-dimensional square arrays, the addition of vorticity can stabilize some low-power modes that are otherwise, unstable \cite{Malomed2001}.

In order to test this idea in conditions that are closer to an actual experiment, we simulate next the beam evolution in our structure by solving the {\em continuous} 2D nonlinear Schr\"odinger equation for the slowly varying electric field envelope $E$: 

\begin{equation}
i\frac{\partial E}{\partial z}+D\left(\frac{\partial^2}{\partial x^2}+\frac{\partial^2}{\partial y^2}\right)E+\frac{2\pi\gamma}{\lambda}|E|^2 E+\frac{2\pi}{\lambda}G(x,y)=0
\end{equation}

Where $\nabla^2_\perp= \partial^2_x+\partial^2_y$ is the transverse Laplacian, $D=\lambda/4\pi n$ is the diffraction coefficient, $\gamma$ is the nonlinear coefficient, $\lambda$ is the wavelength of light, $n$ is the background refractive index, and $G(x,y)$ is the refractive index profile defined numerically as a hexagonal lattice of circular holes with a diameter of $d$, pitch $\Lambda$  which is the distance between the centre of two adjacent holes, and refractive index contrast $\Delta n$. While the general description of the thermal nonlinearity is nonlocal \cite{Rasmussen2009}, here we use the simpler approximation of Kerr type nonlinearity. 

Such model is commonly used to investigate guidance properties in periodic arrays \cite{Christodoulides1988}. Using a hole diameter of $d=5 \mu$m and pitch $\Lambda=10 \mu$m, closely matching a commercial fibre (F-SM 15 by Newport) we are able to provide a theoretical basis for experimental observation.

\begin{figure}
\centerline{\includegraphics[width=1\columnwidth]{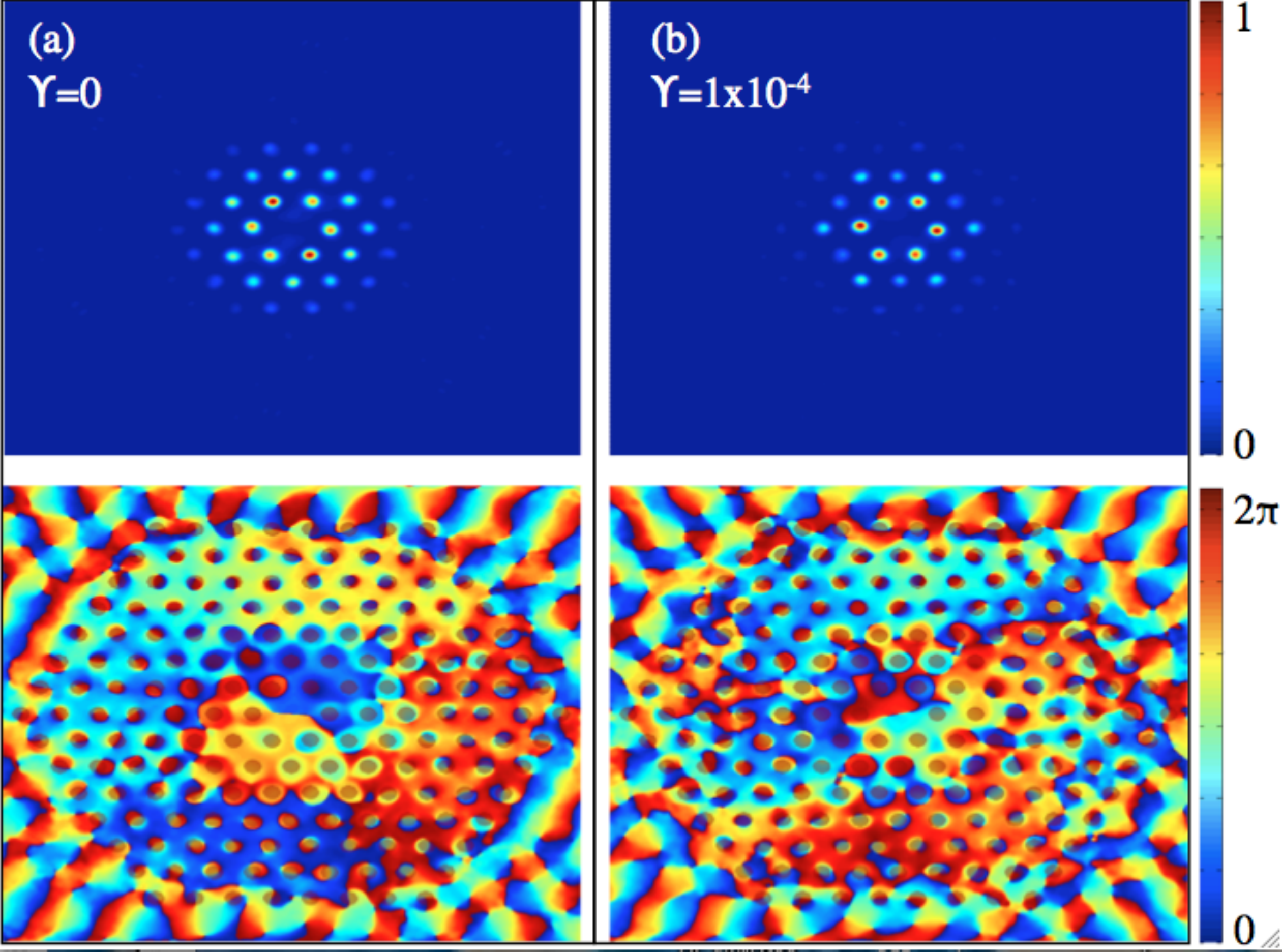}}
\caption{(Color online) (a) Linear propagation of an optical vortex mode with $S=1$ at $z=2$ cm (upper) and its phase (lower) in an hexagonal array of nonlinear waveguides. This linear mode diffracts as it propagates. (b) Nonlinear propagation of the same mode at $z=2$ cm (upper) and its phase (lower) with $\gamma = 1\times 10^{-4}$. The nonlinear mode propagates as a surface mode around a solid core defect. Pale dots on lower plots indicate waveguide location.}
\label{fig6}
\end{figure}

We propagate an input mode with profile $E(z)=r^{|S|}\exp(-\frac{r^2}{w^2}+i S \theta)$, where $E$ is the amplitude of the mode, $S=1$ is the charge of the vortex, $r$, $\theta$ and $z$ are the cylindrical coordinates of the system, and $w=\sqrt{2/|S|}\Lambda$ is the width of the vortex mode.

We find that even though the input mode is symmetric, linear diffraction causes some asymmetry in the output mode after $z=2$ cm propagation (Fig. \ref{fig6}(a)) and index contrast $\Delta n=0.0032$. The vortex phase is somewhat maintained at the output for linear propagation (one can pick a point in the cladding, and trace the phase in a circle around the core from 0 to $2\pi$). This linear beams diffracts in the array as it propagates. Nonlinear output with $\gamma=1\times 10^{-4}$ shows localization of the beam to the first ring of waveguides surrounding the solid core defect (Fig. \ref{fig6}(b)), and the loss of vorticity in the phase. Similar to the discrete model, we see that the vortex mode is unstable when adjacent waveguides are not out of phase. These nonlinear modes are surface modes, largely confined to waveguides around the core for large propagation distance ($z=20$ cm).

\begin{figure}
\centerline{\includegraphics[width=1\columnwidth]{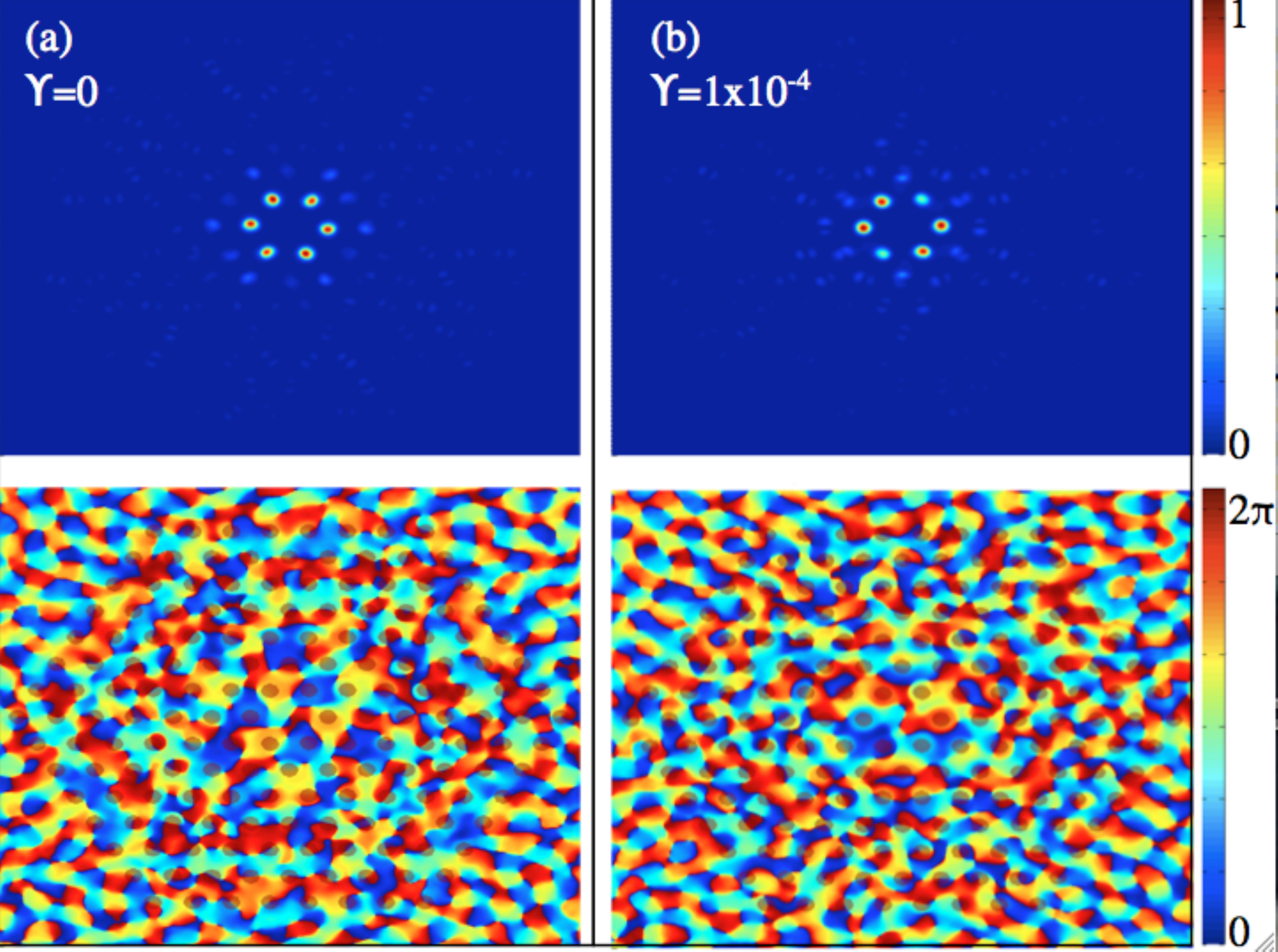}}
\caption{(Color online) (a) Linear propagation of an optical vortex mode with $S=3$ at $z=2$ cm (upper) and its phase (lower) in an hexagonal array of nonlinear waveguides. This linear mode diffracts as it propagates. (b) Nonlinear propagation of the same mode at $z=2$ cm (upper) and its phase (lower) with $\gamma = 1\times 10^{-4}$. The nonlinear mode propagates as a surface mode around a solid core defect. Pale dots on lower plots indicate waveguide location.}
\label{fig7}
\end{figure}

To test the stability in the continuous model we next propagate a beam with $S=3$, which satisfies the condition of staggered phase between adjacent waveguides in the six waveguides surrounding the core. We see linear propagation is indeed more stable, and even more confined in the form of a ring mode, even though the vortex phase is lost (Fig. \ref{fig7}(a)). The linear beam diffracts as it propagates in a similar fashion to $S=1$ modes.  In the nonlinear regime we see the mode begins to break up as the staggered phase is lost (Fig. \ref{fig7}(b)). Again these nonlinear modes are surface modes, although this time more confined to waveguides around the core over large propagation distance ($z=20$ cm), due to the initial staggered phase profile. 

The linear and nonlinear modes produced with this $S=3$ input mode are more strongly confined to the waveguides adjacent to the core defect, when compared to modes produced with an input with $S=1$. While the $\pi$ staggered phase between all six waveguides around the core is lost in the nonlinear propagation, the sites which have the highest intensity maintain this staggered phase. 
The initial vorticity of the $S=3$ beam seems to stabilize the linear output into a vortex which survives for long propagation distances. In the nonlinear regime the vorticity is lost, but the ring mode structure is maintained with a somewhat staggered phase between some waveguides.


\section{Conclusions}
In summary, we have examined the localized surface modes around the core defect of a PCF surrounded by a hexagonal array of nonlinear waveguides. We find that the stable modes in both the discrete and continuous models have a staggered phase profile for the six waveguides surrounding the core. Ring shaped surface modes are studied in the discrete model and shown to always decay to a single site fundamental surface mode. The continuous model shows a similar decay of the surface modes and loss of vorticity in the phase at high nonlinearity. 

It is suggested that this work could be performed in an experimental setting using a liquid infiltrated PCF. One needs to be careful in choosing the parameters of the fibre, and the nonlocal character of the nonlinearity must be taken into account \cite{Rasmussen2009}. 

\section{Acknowledgments}
The authors are grateful to Y. S. Kivshar and D. N. Neshev for useful discussions. M.I.M. acknowledges partial support from FONDECYT Grant  1080374 and Programa de Financiamiento Basal de CONICYT (FB0824/2008).


\begin{thebibliography}{99}

\bibitem{Christodoulides1988}
D.N. Christodoulides and R.I. Joseph, Opt. Lett. {\bf 13}, 794 (1988).

\bibitem{Eisenberg1998}
H.S. Eisenberg and Y. Silberberg, Phys. Rev. Lett. {\bf 81}, 3383 (1998).

\bibitem{Kivshar1993}
Y.S. Kivshar, Opt. Lett. {\bf 18}, 1147 (1993).

\bibitem{Gahagan1999}
K. Gahagan and G. Swartzlander Jr, J. Opt. Soc. Am. B  {\bf 16}, 533 (1999).

\bibitem{Truscott1999}
A. Truscott, M. Friese, N. Heckenberg, and H. Rubinsztein-Dunlop, Phys. Rev. Lett. {\bf 82}, 1438 (1999).

\bibitem{Ferrando2004}
A. Ferrando, M. Zacares, P. Fernandez De Cordoba, D. Binosi, and J.A. Monsoriu, Opt. Express {\bf 12}, 817 (2004).

\bibitem{Johansson1998}
M. Johansson, S. Aubry, Y. B. Gaididei, P. L. Christiansen, K. O. Rasmussen,  Physica D {\bf 119}, 115 (1998).

\bibitem{Kevrekidis2001}
P. G. Kevrekidis, B. A. Malomed, A. R. Bishop, D. J. Frantzeskakis, Phys. Rev. E {\bf 65}, 016605 (2001).

\bibitem{Malomed2001}
B. A. Malomed and P. G. Kevrekidis, Phys. Rev. E {\bf 64}, 026601 (2001).

\bibitem{Kevrekidis2002}
P. G. Kevrekidis, B. A. Malomed, and Yu. B. Gaididei, Phys. Rev. E {\bf 66}, 016609, (2002).

\bibitem{Yang2003}
•J. Yang and Z. Musslimani, Opt. Lett. {\bf 28}, 2094 (2003). 

\bibitem{Alexander2007}
T.J. Alexander, A.S. Desyatnikov, and Y.S. Kivshar, Opt. Lett. {\bf 32}, 1293 (2007).

\bibitem{Neshev2004}
D. Neshev, T. Alexander, and E. Ostrovskaya, YS, Phys. Rev. Lett. {\bf 92},  123903 (2004).

\bibitem{Rasmussen2009}
P.D. Rasmussen, F.H. Bennet, D.N. Neshev, A. a Sukhorukov, C.R. Rosberg, W. Krolikowski, O. Bang, and Y.S. Kivshar, Opt. Lett. {\bf 34}, 295 (2009).

\bibitem{Bennet2010}
F.H. Bennet, I.A. Amuli, A.A. Sukhorukov, W. Krolikowski, D.N. Neshev, and Y.S. Kivshar,  Opt. Lett. {\bf 35}, 3213 (2010).

\bibitem{Bennet2010a}
F.H. Bennet and J. Farnell,  Opt. Comm. {\bf 283}, 4069 (2010).

\bibitem{Wu2009}
D. Wu, B. Kuhlmey, and B. Eggleton, Opt. Lett. {\bf 34}, 322 (2009).

\bibitem{Vieweg2010}
M. Vieweg, T. Gissibl, S. Pricking, B. Kuhlmey, D. Wu, B. Eggleton, and H. Giessen, Opt. Express, {\bf 18}, 25232 (2010).

\bibitem{Szameit2009}
A. Szameit, Y. Kartashov, M. Heinrich, and F. Dreisow, T,  Opt. Lett. {\bf 34}, 797 (2009).

\bibitem{Szameit2008}
A. Szameit, Y. Kartashov, V. Vysloukh, and M. Heinrich, Opt. Lett. {\bf 33}, 1542 (2008).

\bibitem{Molina2006}
M. I. Molina, R. A. Vicencio and Y. S. Kivshar, Opt. Lett. {\bf 31}, 1693 (2006).





\end{thebibliography}
\end{document}